\documentclass[hyper,a4paper,11pt,oneside]{JHEP3}

\JHEPspecialurl{http://jcap.sissa.it/JOURNAL/JHEP3.tar.gz}

\usepackage{graphicx}
\usepackage{dcolumn}
\usepackage{bm}
\usepackage{epsfig,multicol,bbm}

\newcommand\fverb{\setbox\fverbbox=\hbox\bgroup\verb}
\newcommand\fverbdo{\egroup\medskip\noindent%
            \fbox{\unhbox\fverbbox}\ }
\newcommand\fverbit{\egroup\item[\fbox{\unhbox\fverbbox}]}
\newbox\fverbbox

\newcommand{\M}{{\cal {M}}}
\newcommand{\B}{{\cal {B}}}

\newcommand{\HH}{{\cal {H}}}

\title{Dynamics of a self--gravitating neutron source.}
\author{D. Manreza Paret$^{\clubsuit}$, A. P\'{e}rez Mart\'{\i}nez$^{\diamondsuit}$, A. Ulacia Rey$^{\diamondsuit,\spadesuit}$
and Roberto A. Sussman.$^\spadesuit$\\$^{\clubsuit}$Departamento de
F\'{\i}sica General, Facultad de F\'{\i}sica, Universidad de La
Habana,\\San L\'azaro y L, cp-10400, La Habana, Cuba.
\\\email{dmanreza@fisica.uh.cu}\\\\
$^{\diamondsuit}$Departamento de F\'{\i}sica Te\'{o}rica.
\\Instituto de Cibern\'etica, Matem\'atica y F\'{\i}sica, ICIMAF.\\
Calle E No-309 Vedado, cp-10400. La Habana, Cuba.\\\email{aurora,alain@icmf.inf.cu}\\\\
$^{\spadesuit}${Departamento de Gravitaci\'{o}n y Teor\'{\i}as de
Campo.\\ Instituto de Ciencias Nucleares, ICN, \\
Universidad Aut\'{o}noma de M\'{e}xico UNAM, Mexico,DF.
04510.}\\\email{sussman@nucleares.unam.mx}}

\received{18/December/2009}       
\revised{17/February/2010}
\accepted{18/February/2010}       

\preprint{\grqc{0812.2508}}  

\abstract{We examine the dynamics of a self--gravitating magnetized
neutron gas as a source of a  Bianchi I spacetime described by the
Kasner metric. The set of Einstein-Maxwell field equations can be
expressed as a dynamical system in a 4-dimensional phase space.
Numerical solutions of this system reveal the emergence of a
point--like singularity as the final evolution state for a large
class of physically motivated initial conditions. Besides the
theoretical interest of studying this source in a fully general
relativistic context, the resulting idealized model could be helpful
in understanding the collapse of local volume elements of a neutron
gas in the critical conditions that would prevail in the center of a
compact object.}


\begin{document}
\section{Introduction}
\label{intro}

Critical stellar configurations, such as white dwarfs, neutron
stars, supermassive stars, relativistic star clusters and black
holes are important astrophysical systems in which relativistic
effects cannot be ignored. These astrophysical systems, denoted
generically by the term ``compact objects'', can be thought of as
natural laboratories to understand a wealth of phenomena relevant to
theoretical and experimental physics under critical conditions.
These conditions are ideally suited to propose and test theoretical
models of strong magnetic fields associated with self--gravitating
sources under critical conditions. In such conditions, gravitation
couples with other fundamental interactions (strong, weak and
electromagnetic) and from this interplay important clues of their
unification could emerge.

The presence and effects of strong magnetic fields in compact
objects has been studied  in the literature (see
\cite{shapiro,bocquet,Peng:2007uu,Reisenegger:2008et,Suh:2000ni,Cardall},
and references quoted therein). At a very basic level, the simplest
approach is to consider various types of self--gravitating and
magnetized plasmas of neutron matter under equations of state that
are appropriate for the critical conditions of a compact object
\cite{salgado1,salgado2}. Since astrophysical systems exhibit, in
general, angular momentum and pressure anisotropies, it is
interesting to examine how these effects could influence their
stability.

Following the basic known methodology \cite{shapiro}, an of equation
of state based on a nuclear ferromagnetic model (not related to
electric currents) was examined in \cite{Aurora1}, with the purpose
of studying the interplay of pressure anisotropy and the magnetic
field within a Newtonian framework. Hence, in the present work we
consider a relativistic generalization that will allows us to
examine (under specific restriction of the geometry) the evolution
of this type of magnetized neutron gas under strong gravity. In
particular, our aim is to provide the minimally basic qualitative
and quantitative elements to address the question of whether the
magnetic field can ``slow down'', or even reverse or stop,
gravitational collapse, as well as the issue of the evolution to a
stable configuration.

It is evident that a proper and comprehensive study of a magnetized
fermion gas, as a source of a compact object, requires a spacetime
with axial symmetry (or without symmetries), leading to dynamical
equations that must be solved by means of hydrodynamic codes of high
complexity \cite{bocquet,Hartle1,Hartle2,Bednarek}. In previous
articles \cite{Alain2,Ulacia Rey:2006hx} we studied a
self--gravitating gas of magnetized electrons, described by an
appropriate equation of state, as the source of an anisotropic
Bianchi-I spacetime described by the Kasner metric. While a Bianchi
I model is obviously inadequate as the metric of a compact object,
it is among the less complicated geometries compatible with a
magnetized source.  Thus, the evolution of such a source under a
much simplified Bianchi geometry is basically a toy model, but as
such it can still be useful to obtain qualitative features of the
sources under local critical conditions, specially in conditions
near the center of the object where angular momentum plays a minor
dynamical role. These qualitative results could provide a better
understanding, and/or a useful approximated description, of the
behavior of local internal volume elements near the center of a more
realistic configuration. In this article we extend previous work on
the electron gas to the case of a neutron gas. Hence, we follow a
similar methodology based on re--writing the field and conservation
equations for the magnetized neutron gas in the Bianchi I geometry
as a dynamical system, evolving in a 4--dimensional phase space.
This system is then analyzed qualitatively and numerically.

It is worthwhile remarking the basic differences between the
magnetized electron gas examined in \cite{shapiro,Alain2,Ulacia
Rey:2006hx} and the neutron gas that we consider in this article.
Electrons interact with a magnetic field through their charge,
leading to the Landau diamagnetism characterized by a quantization
effect associated with the Landau energy levels. However, neutrons
interact with the magnetic field by their anomalous magnetic moment
(AMM), in the context of Pauli's paramagnetism and the equations of
Pauli--Dirac. Consequently, one expects the magnetic neutron
interaction to be weaker, though conditions of degenerated neutron
gases in compact objects are also expected to be more critical
(because of the higher densities) than those of degenerated electron
gases. Therefore, relativistic effects of gravity are more likely to
play a dominant dynamical role in neutron gases. Still, it is
important to mention that a self--gravitating and magnetized neutron
gas is a simplified model of a source for a compact object, as
protons and chemical equilibrium potentials should also be
considered to evaluate local
effects~\cite{shapiro,salgado1,salgado2}. Nevertheless, the
magnetized neutron gas already exhibits important qualitative and
quantitative differences in comparison with electron gases
previously examined.

 The paper is organized as follows:  we derive in section II the equation of
state for a magnetized neutron gas. The Einstein--Maxwell field
equations for the Kasner metric and this source are derived in
section III. In section IV we examine the limit of a weak magnetic
field. The set of ordinary non--linear differential equations that
follows from the field equations is transformed into a set of
evolution equations in sections V and VI. This dynamical system is
analyzed qualitatively and numerically in section VII. Our
conclusions are presented in sections VIII.
\section{Magnetized neutron gas: the equation of state.}
\label{sec:1}

 The main properties of gas of magnetized degenerated neutrons are well
known (see \cite{shapiro,Aurora1,Guang}). Considering the grand
canonical ensemble, a subsystem can be thought of as a local volume
of the neutron gas under the influence of a magnetic field $\vec{H}$
associated with the rest of the system (in an astrophysical context
this could be good approximation to a suitable volume element inside
a compact object). Because of this field, the subsystem becomes
polarized, leading to a magnetization vector that satisfies the
relationship:\, $\vec{H}=\vec{\B}-4\pi \vec{\M}$. The field
$\vec{H}$ can be though of as ``external'' to the subsystem, while
$\vec{\B}$ can be ``internal'' to any particle within the subsystem,
which feels (in addition to $\vec{H}$) the contribution
$4\pi\vec{\M}$ from particles from the rest of the subsystem.

The equation of state of the neutron gas follows from calculating
the energy spectrum of the particles that make up the system. We can
obtain this energy spectrum from the Dirac equation for neutral
particles with anomalous magnetic moment:
\begin{equation}\label{Diraceq}
(\gamma^{\mu}\partial_{\mu}+m+iq\sigma_{\mu \nu}F^{\mu \nu})\Psi=0,
\end{equation}
 where $\sigma_{\mu\lambda}=\frac{1}{2}(\gamma_\mu\gamma
_\lambda-\gamma _\lambda\gamma _\mu)$ is the spin tensor,\, $F^{\mu
\nu}$ is the electromagnetic field tensor (we have set
$\hbar=\mathrm{c}=1$) and $\Psi$ is the Dirac field. Solving
equation (\ref{Diraceq}) leads to the following energy spectrum
\cite{Guang}, \cite{Bagrov},\cite{Wen:2005kf}:
\begin{equation}
E_n(p,\B,\eta)=\sqrt{p_{\|}^2+(\sqrt{p_\bot^2+m_n^2}+\eta q\B)^2},
\end{equation}
where $p_{\|}, p_\bot$ are, respectively, the components of momentum
in the directions parallel and perpendicular to the magnetic field
$\B$, $m_n$ is the mass of the neutron, $q=-1.91\mu_N$ is the
neutron magnetic moment ($\mu_N=e/2m_p$ is the nuclear magneton),
$\eta=\pm1$ are $\sigma_3$ eigenvalues corresponding to the two
possible orientations (parallel or antiparallel) of the neutron
magnetic moment with respect to the magnetic field.

The thermodynamical grand potential then takes the form
\begin{equation}
\Omega=-kT \ln\mathcal{Z},
\end{equation}
where $k$ is the Boltzman constant, $T$ is the temperature,
$\mathcal{Z}=Tr(\hat{\rho})$ is the partition function of the
system, $\rho = e^{-(\hat{H}-\mu \hat{N})/kT}$, \, $\hat{H}$ is the
Hamiltonian, $\mu$ is the chemical potential and $\hat{N}$ is the
 number of particles operator.

The energy--momentum tensor associated with an external constant
magnetic field takes the form:
\begin{equation}\label{tensor_e-m}
\mathcal{T}^{\mu}\,_{\nu}=(T\frac{\partial{\Omega}}{\partial{T}}+\sum{\mu_{n}\frac{\partial{\Omega}}{
\partial{\mu_{n}}}})\delta^{\mu}_{\,\,4}\,\delta^{4}_{\,\,\nu}+4F^{\mu \gamma}F_{\gamma \nu}\frac{\partial{\Omega}}{
\partial{F^{2}}}-\Omega\,\delta^{\mu}\,_{\nu},
\end{equation}
so that in the limit of zero magnetic field limit we obtain the
perfect fluid tensor:\,
$\mathcal{T}^{\mu}\,_{\nu}=p\delta^{\mu}\,_{\nu}-(p+U)\delta^{\mu}_{\,\,4}\delta^{4}_{\,\,\nu}$.
The components of the tensor (\ref{tensor_e-m}) are:
\begin{eqnarray}
\mathcal{T}^{3}_{\,\,\,\,3}&=&=p_{\|}=-\Omega=p,
\\
\mathcal{T}^{1}_{\,\,\,\,1}&=&\mathcal{T}^{2}_{\,\,\,\,2}=p_{\perp}=-\Omega-\B\M=p-\B\M,
\\
\mathcal{T}^{4}_{\,\,\,\,4}&=&-U=-TS-\mu N-\Omega,
\end{eqnarray}
\label{componentes de T}
%
where $S$ is the specific entropy, $N=-\partial\Omega/\partial\mu$
is the particle number density, $\M=-\partial\Omega/\partial\B$ is
the magnetization, $U$ is the energy density and $p$ is the pressure
in the direction of the magnetic field.

The thermodynamical potential can be split in the following two
parts:
\[
\Omega=\Omega_{sn}+\Omega_{Vn},
\]
where the first term in the right hand side is the statistical
contribution and the second is the vacuum contribution
\cite{Aurora1}.  Explicitly, we have
\begin{equation}\label{omega_n}
\Omega_{sn}=-\frac{1}{4\pi^2\xi}\sum_{\eta=1,-1}^{}\int_0^\infty
p_\bot \mathrm{d}p_\bot \mathrm{d}p_3
\mathrm{ln}[f^+(\mu_n,\xi)f^-(\mu_n,\xi)],
\end{equation}
where $\xi=1/k_BT$ and $f^{\pm}(\mu_n,\xi)=(1+e^{(E_n\mp\mu_n)\xi})$
represent, respectively, the contributions from the particles and
from the antiparticles. The vacuum term is given by the expression:
\begin{equation}\label{omega_v}
\Omega_{Vn}=\frac{1}{4\pi^2\xi}\sum_{\eta=1,-1}^{}\int_0^\infty
p_\bot \mathrm{d}p_\bot \mathrm{d}p_3 E_n,
\end{equation}
which is divergent, but can be renormalized, and for fields of
intensity $\B<10^{18}\,\mathrm{G}$ its contribution is not
important~\cite{Aurora1}, hence we will neglect this term in the
remaining of the present article.

Equation (\ref{omega_n}) can be easily integrated in the case that
concerns us $(T=0)$, leading to
\begin{equation}\label{omega}
\Omega_{sn}=-\lambda\sum_{\eta=1,-1}\biggl[\frac{\mu f_\eta^3}{12}+
\frac{(1+\eta \beta)(5\eta \beta-3)\mu f_\eta}{24}+\frac{(1+\eta
\beta)^3(3-\eta \beta)}{24}L_\eta-\frac{\eta \beta
\mu^3}{6}s_\eta\biggr],
\end{equation}
where we have introduced the following expressions:
\begin{equation}
f_\eta=\sqrt{\mu^2-(1+\eta\beta)^2}, \quad
s_\eta=\frac{\pi}{2}-\arcsin\biggr(\frac{1+\eta\beta}{\mu}\biggl),
\quad \mu=\frac{\mu_n}{m_n} \label{efe}
\end{equation}
\begin{equation}
L_\eta=\ln\biggl(\frac{\mu+f_\eta}{1+\eta\beta}\biggr), \quad
\beta=\frac{\B}{\B_c},
\end{equation}
 with $\B_c=m_n/q \simeq
1.56\times10^{20}\,\mathrm{G}$ being the critical field and
$\lambda=m_n^4/4\pi^2\hbar^3\mathrm{c}^3=4.11\times10^{36}\mathrm{\,erg
\,cm^{-3}}$.

All thermodynamical variables of the system follow readily from the
thermodynamical potential $\Omega$. In particular, by computing
neutron density and magnetization in this way we obtain
$N=N_0\Gamma_N$, $\M=\M_0\Gamma_M$, where $N_0=\lambda/m_n$,
$\M_0=N_0q$, while the coefficients $\Gamma_N,\Gamma_M$ take the
form:

\begin{eqnarray*}
\Gamma_N&=& \sum_{\eta=1,-1}^{}\biggl[\frac{f_\eta^3}{3}+\frac{\eta
\beta(1+\eta\beta)f_\eta}{2}-\frac{\eta\beta \mu^2}{2}s_\eta\biggr],\\
\Gamma_M&=&-\sum_{\eta=1,-1}^{}\eta\biggl[\frac{(1-2\eta \beta)\mu
f_\eta}{6}-\frac{(1+\eta\beta)^2(1-\eta\beta/2)}{3}L_\eta+\frac{\mu^3}{6}s_\eta
\biggr].
\end{eqnarray*}\label{gamma1}

Therefore, given (\ref{componentes de T}) and (\ref{omega}), we can
write the equation of state for a relativistic degenerated neutron
gas in the presence of an external magnetic field as:

\begin{eqnarray}
U&=&\mu_n N+\Omega=\lambda \Gamma_U(\beta,\mu),\label{EOS1}
\\
p&=&-\Omega=\lambda \Gamma_P(\beta,\mu),\label{EOS2}
\\
M&=&\B \M=\lambda\beta \Gamma_M(\beta,\mu),\label{EOS3}
\end{eqnarray}
\label{EOS}

where

\begin{eqnarray*}
\Gamma_P&=&\sum_{\eta=1,-1}^{}\biggl[\frac{\mu f_\eta^3}{12}+
\frac{(1+\eta \beta)(5\eta \beta-3)\mu f_\eta}{24}+
\\
&&\hspace{3cm}+\frac{(1+\eta\beta)^3(3-\eta\beta)}{24}L_\eta-\frac{\eta
\beta \mu^3}{6}s_\eta\biggr],
\\
\Gamma_U&=&\mu \Gamma_N-\Gamma_P.
\end{eqnarray*}
%

Let us remark at this point that in (\ref{omega})-(\ref{EOS3}) we
are summing over the magnetic moments parallel ($\eta=-1$) or
antiparallel ($\eta=1$) to the magnetic field ({\it{i.e.}} the well
known Pauli Paramagnetism). The choice $\eta=\pm 1$, \cite{Aurora1}
is equivalent to consider different phases  of the system. The
appearance of the threshold of the value of the magnetic field for
each one of these cases can be seen if we analyze the expressions of
the functions $f_{\eta}$ and $s_{\eta}$ in (\ref{efe}). We have take
$\beta\geq 0$ therefore starting with $f_{\eta}$:

\begin{equation}
\mu^2\geq (1+\eta \,\beta)^2=1+2\eta\beta+\eta^2\beta^2 .
\end{equation}
however as $\eta=-1,1$ then always $\eta^2=1$ and we can rewrite:
\begin{equation}
\mu^2\geq (\eta+\beta)^2.
\end{equation}
hence;
\begin{equation}
(\beta+\eta-\mu)(\beta+\eta+\mu)\leq 0.
\end{equation}

%
%
%
%
Now we have two possibilities. But, it is easy to realize that the
only acceptable possibility is:
\begin{equation}
-\mu-\eta\leq\,\beta\,\leq \mu-\eta.  \label{CasoB}
\end{equation}
This is exactly the restriction that comes from the function
$s_{\eta}$:
\begin{equation}
\mid \frac{1+\eta\beta}{\mu}\mid \leq 1.  \label{arcsin_restriccion}
\end{equation}
Therefore, from either one (\ref{CasoB}) or
(\ref{arcsin_restriccion}) the following constraints are obtained
for the magnetic field:
\begin{eqnarray}
if \quad \eta=1 \,  \Rightarrow \qquad -1-\mu \leq \beta \leq \mu-1,  \\
if \quad \eta=-1 \, \Rightarrow \qquad 1-\mu \leq \beta \leq 1+\mu.
\end{eqnarray}
These last inequalities restrict the values of the magnetic field.
Thus neutrons with AMM aligned to the magnetic field have only
magnetic field values in the range between $1-\mu \leq \beta \leq
1+\mu$. Similarly, for neutrons with AMM oriented antiparallel to
the magnetic field can take magnetic field values
 in the interval $-1-\mu \leq \beta \leq \mu-1$. In
the  particular case $\mu=1$ we have.
\begin{eqnarray}
   if \quad \eta=-1, \qquad \, 0 \leq \beta \leq 2,\\
   if \quad \eta=1,  \qquad \, -2 \leq \beta \leq 0.
\end{eqnarray}
This means that for fields smaller  (or equal) than two times the
critical magnetic field, the neutrons are aligned with $\eta=-1$
(the AMM is parallel to the magnetic field). Hence, neutrons with
$\eta=1$ are forced to invert its sense.
\section{Einstein--Maxwell equations.}
\label{sec:2} If a local self--gravitating volume of a magnetized
neutron gas evolves in the conditions prevailing inside a compact
object, relativistic effects will be important. This implies that
local dynamics must be studied in the framework of General
Relativity, by means of Einstein's field equations:
\begin{equation}
G_{\mu\nu}=R_{{\mu\nu}}-\frac{1}{2}R\, g_{{\mu\nu}}=\kappa\,
\mathcal{T}_{\mu\nu}. \label{EE1}
\end{equation}
together with energy--momentum balance and Maxwell equations:
\begin{eqnarray} \label{Ebal}\mathcal{T}^{\mu\nu}\,_{;\,\nu}=0,\\
\label{Maxwell_eq} F^{\mu\nu}\,_{;\,\nu}=0, \qquad
F_{[\,\mu\nu\,;\,\alpha\,]}=0.\end{eqnarray}
where $\kappa=8\pi \mathrm{G_N}$, with $\mathrm{G_N}$ being
Newtons's gravitational constant, while square brackets denote
anti--symmetrization in $\mu\nu;\alpha$. The energy--momentum tensor
$\mathcal{T}^{\mu}_{\,\,\,\,\nu}$ associated to the magnetized
neutron gas is given by (\ref{componentes de T}), with the relevant
thermodynamical potentials obtained through Statistical Mechanics
and complying with the appropriate equation of state (as discussed
in the previous section). This tensor can also be given  in terms of
the 4-velocity $u^{\alpha}$ field as:
\begin{equation}
\mathcal{T}^{\alpha}\,_{\beta}=(U+\widetilde{P})u^{\alpha}u_{\beta}+\widetilde{P}\,\delta^{\alpha}\,_{\beta}+\Pi^{\alpha}\,_{\beta},
\qquad \widetilde{P}=p-\frac{2\B\M}{3}. \label{TE-M}
\end{equation}

We will consider the field equations (\ref{EE1})--(\ref{Maxwell_eq})
with (\ref{TE-M}) as the source tensor of a Bianchi I model
described in the coordinate representation known as the Kasner
metric:
\begin{equation}
{ds^2}=Q_{1}(t)^2dx^{2}+Q_{2}(t)^2dy^2+Q_{3}(t)^2 d{z}^2-dt^2.
\label{Metrica-K}
\end{equation}
which suggests considering a comoving geodesic 4--velocity
$u^\alpha=\delta^\alpha_t$, so that the anisotropic pressure tensor
in (\ref{TE-M}) in the coordinates $[x,y,z,t]$ takes the form:
\begin{equation}
\Pi^{\alpha}\,_{\beta}=\hbox{diag}\,[\Pi,\Pi,-2\Pi,0], \qquad
\Pi=-\frac{\B\M}{3},\qquad \Pi^{\alpha}\,_{\alpha}=0. \label{TAn}
\end{equation}
The field equations (\ref{EE1}) for (\ref{TE-M}) and (\ref{TAn})
take the form:

\begin{eqnarray}\label{EE2}
-G^{x}\,_{x}&=&\frac{\dot{Q_{2}}\dot{Q_{3}}}{Q_{2}Q_{3}}+\frac{\ddot{Q_{2}}}{Q_{2}}+\frac{\ddot{Q_{3}}}{Q_{3}}=-\kappa(p-\B\M),
\\
-G^{y}\,_{y}&=&\frac{\dot{Q_{1}}\dot{Q_{3}}}{Q_{1}Q_{3}}+\frac{\ddot{Q_{1}}}{Q_{1}}+\frac{\ddot{Q_{3}}}{Q_{3}}=-\kappa(p-\B\M),
\\
-G^{z}\,_{z}&=&\frac{\dot{Q_{1}}\dot{Q_{2}}}{Q_{1}Q_{2}}+\frac{\ddot{Q_{1}}}{Q_{1}}+\frac{\ddot{Q_{2}}}{Q_{2}}=-\kappa
p,
\\
-G^{t}\,_{t}&=&\frac{\dot{Q_{1}}\dot{Q_{2}}}{Q_{1}Q_{2}}+\frac{\dot{Q_{1}}\dot{Q_{3}}}{Q_{1}Q_{3}}+\frac{\dot{Q_{2}}\dot{Q_{3}}}{Q_{2}Q_{3}}=\kappa
U. \label{EE2_Gtt}
\end{eqnarray}

where $\dot A= A_{;\alpha} u^{\alpha}=A_{,t}$. Energy--momentum
balance (\ref{Ebal}) leads to:
\begin{equation}\label{eq_U[t]}
\dot{U}=\frac{\dot{Q_{3}}}{Q_{3}}(p+U)-(\frac{\dot{Q_{1}}}{Q_{1}}+\frac{\dot{Q_{2}}}{Q_{2}})(-\B\M+p+U).
\end{equation}
while the Maxwell equations equations (\ref{Maxwell_eq}) imply:
\begin{equation}\label{Maxwell_eq_1}
\frac{\dot{Q_{1}}}{Q_{1}}+\frac{\dot{Q_{2}}}{Q_{2}}+\frac{1}{2}\frac{\dot{\B}}{\B}=0.
\end{equation}
The Einstein-Maxwell equations (\ref{EE2}, \ref{eq_U[t]} and
\ref{Maxwell_eq_1}) are non--linear second order ordinary
differential equations for the metric functions $Q_{1}, Q_{2},
Q_{3}$ and $U$. In order to treat this system numerically, it is
necessary to
 introduce a new set of variables that will transform it into a system of
first order evolution equations. However, before undertaking this
task,  we examine in the following section the weak field limit.
\section{Limit of weak magnetic field.}
The discussion of the limit of weak magnetic field is important in
to illustrate the connection between our quantum magnetic field and
a classic Maxwellian field in the context of a magnetohydrodynamic
treatment. In sections 2 we have shown equations strongly linked
with quantum magnetic field.

 It is important to emphasize that the term ``quantum magnetic
field'' involves the semi--classical interaction between the
magnetic field and the anomalous magnetic moment. This approach
implies a theoretical connection between the equation of state
introduced in the previous section and a QED framework. As a
consistency condition, this framework must allow for a classical
Maxwellian limit that should arise from a series expansion around
the zero of the magnetic field $\beta=0$. The leading term of this
expansion should lead to the well known energy-momentum tensor for a
Maxwellian magnetic field \cite{MTWGravitation}. In general, this
type of series expansions can be done as follows:

\begin{eqnarray}
p_{\perp}&=& \sum_{n=0}^{\infty}
(\frac{\partial^{n}{p_{\perp}}}{\partial{\beta^n}})\mid_{_{\beta=0}}\frac{\beta^{n}}{n!}\simeq
p_{1}-a_{1}\beta^2+\mathcal{O}(\beta^4),\label{WMF1}\\
p_{\|}&=& \sum_{n=0}^{\infty}
(\frac{\partial^{n}{p_{\|}}}{\partial{\beta^n}})\mid_{_{\beta=0}}\frac{\beta^{n}}{n!}\simeq
p_{1}+a_{3}\beta^2+\mathcal{O}(\beta^4),\label{WMF2}
\\
U&=& \sum_{n=0}^{\infty}
(\frac{\partial^{n}U}{\partial{\beta^n}})\mid_{_{\beta=0}}\frac{\beta^{n}}{n!}\simeq
U_{0}+a_{o}\beta^2+\mathcal{O}(\beta^4), \label{WMF3}
\end{eqnarray}
where it is easy to see that: $p_{1}=p_{\|}(\beta=0)\equiv
p_{\perp}(\beta=0),\,a_{0}=({\partial^{2}U}/{\partial{\beta^2}})\mid_{_{\beta=0}}/2
,\,a_{1}=({\partial^{2}{p_{\perp}}}/{\partial{\beta^2}})\mid_{_{\beta=0}}/2
,\,a_{3}=({\partial^{2}{p_{\|}}}/{\partial{\beta^2}})\mid_{_{\beta=0}}/2,\,
U_{0}=U(\beta=0)$. Hence, all of the former are functions only of
the dimensionless chemical potential
 $\mu$.

As shown in the examples in the literature of classical magnetic
fields in the context of a Bianchi-I
geometry~\cite{Tsagas:1999tu,Barrow:2006ch}, the leading magnetic
field terms are quadratic. This suggest to truncate the series in
(\ref{WMF1})-(\ref{WMF3}) in the quadratic terms, as the higher
order terms, like $\beta^4,\,\beta^6,...$, are (in general)
extremely small multipole contributions. If we assume that the
velocity fluctuations of the plasma tend to a zero average
macroscopically, and that the medium does not undergo any bulk
motion, then these contributions can be safely neglected (though
typically higher velocities could arise from thermal fluctuations or
quantum disorder). Under these assumptions, the energy-momentum
tensor of a neutron gas  with a minimally coupled magnetic field can
always be written in the form \cite{Tsagas:1999tu}:
\begin{equation}
T_{\mu \nu}=(U_{0}+U_{mag})u_{\mu}u_{\nu}+(p_{0}+p_{mag})h_{\mu
\nu}+\Pi^{mag}_{\mu \nu}.\label{T_Roy}
\end{equation}
where $p_{mag}$ and $U_{mag}$ are, respectively, the magnetic
pressure and the magnetic energy.

 It is important to remark that $p_{0}$ in (\ref{T_Roy}) is the
isotropic contribution to the pressure of the system, which (in
general) can depend on the chemical potential and the magnetic
field. On the other hand, $p_{1}$ is the pressure for the case
without magnetic field $\beta=0$. In the case under consideration we
have:
\begin{equation}
T^{\mu}\,_{\nu}=\textrm{diag}[p_{1}-a_{1}\beta^2,p_{1}-a_{1}\beta^2,p_{1}+a_{3}\beta^2,-U_{0}-a_{0}\beta^2].\label{T_AUR_DEBIL}
\end{equation}
Comparison with equations (\ref{T_AUR_DEBIL})-(\ref{T_Roy}) leads
to:
\begin{eqnarray}
p_{mag}&=&\frac{H^{2}}{6}=a_{0}\beta^2/3,
\\
U_{mag}&=&\frac{H^{2}}{2}=a_{0}\beta^2.
\end{eqnarray}
where $H^2=H^{\mu}H_{\mu}=2a_{0}\beta^2$ and $H_{\mu}$ are the
components of the magnetic field, which we have assumed to point in
the $z$--direction. This assumption is consistent with the fact that
a small volume element around the center of a compact object is
approximately homogeneous and the rotation axis furnishes a
privileged direction, which we can always align with the $z$ axis:
\begin{equation}
H^{\alpha}=(0,0,\sqrt{2a_{0}}\frac{\beta}{Q_{3}},0), \qquad
\,H_{\alpha}=(0,0,\sqrt{2a_{0}}\beta Q_{3},0) .
\end{equation}
The tensor $\Pi^{mag}_{\mu \nu}$ is then the projected symmetric
trace-free tensor representing anisotropic pressures that comes from
the magnetic field. It can be written as:
\begin{equation}
\hspace{0.8cm}(\Pi^{mag})^{\mu}_{\,\,\nu}=\textrm{diag}[-\frac{1}{3}(a_{1}+a_{3})\beta^{2},-\frac{1}{3}(a_{1}+
a_{3})\beta^{2},\frac{2}{3}(a_{1}+a_{3})\beta^{2},0].\label{PI_mag}
\end{equation}
Is important to point out that $p_{1}$ in (\ref{T_AUR_DEBIL}) is, in
general, different from $p_{0}$. Thus:
\begin{equation}
p_{0}=p_{1}-(a_{0}+2a_{1}-a_{3})\frac{\beta^2}{3}.
\end{equation}
where $p_{1}=p_{(\beta=0)}=\widetilde{P}_{(\beta=0)}$ in
(\ref{EOS2}) and (\ref{TE-M}) is the pressure without magnetic
field. Only when the magnetic field is zero, $\beta=0$, the
pressures coincide: $p_{0}=p_{1}$ and the energy--momentum tensor
becomes that of a perfect fluid with isotropic pressure. In this
case  $p_{0}$ and $U_{0}$ correspond to the pressure and energy
density of a classical neutron gas. \\
\section{Local kinematic variables.}

Since we are interested on the local evolution of volume elements of
the magnetized neutron gas associated with the source (\ref{TE-M}),
we need to re--write the dynamical Einstein--Maxwell equations in
terms of covariant parameters associated with the local kinematics
of volume elements as described by $u^{\alpha}$. For the Kasner
metric in the comoving geodesic 4--velocity frame, the nonzero local
kinematic parameters are the expansion scalar, $\Theta$, and shear
tensor, $\sigma^{\alpha \beta}$, given by:
\begin{eqnarray} \label{expsc}\Theta &=& u^{\alpha}\,_{;\alpha}\\
\label{shear} \sigma_{\alpha\beta} &=&
u_{(\alpha;\beta)}-\frac{\Theta}{3}\,h_{\alpha\beta}.\end{eqnarray}
where $h_{\alpha\beta}=u_{\alpha}\,u_{\beta}+g_{\alpha \beta}$ is
the projection tensor and round brackets denote symmetrization on
the indices $\alpha,\beta$. The geometric interpretation of these
parameters is straightforward: $\Theta$ denotes the isotropic proper
time change of proper volume of local fluid elements, whereas
$\sigma^{\alpha}\,_{\alpha}$ describes the deformation of local
volumes as they expand at different rates along the directions given
by its eigenvectors.

The expansion scalar and components of the shear tensor for the
Kasner metric are:
\begin{eqnarray}\label{ec_teta_K}
\Theta &=& \frac{\dot{Q_{1}}}{Q_{1}}+\frac{\dot{Q_{2}}}{Q_{2}}+\frac{\dot{Q_{3}}}{Q_{3}},\\
\sigma^\mu\,_\nu &=&
\hbox{diag}\,[\sigma^x\,_x,\,\sigma^y\,_y,\,\sigma^z\,_z,\,0] =
\hbox{diag}\,[\Sigma_1,\,\Sigma_2,\,\Sigma_3,\,0],\end{eqnarray}
where
\begin{equation}
\Sigma_{a}=\frac{2\dot{Q_{a}}}{3Q_{a}}-\frac{\dot{Q_{b}}}{3Q_{b}}-\frac{\dot{Q_{c}}}{3Q_{c}},
\qquad a \neq b \neq c,\, (a,b,c=1,2,3). \label{sigma_comp}
\end{equation}
Since the shear tensor is trace--free:
$\sigma^{\alpha}\,_{\alpha}=0$, we can eliminate one of the
quantities $(\Sigma_1,\,\Sigma_2,\,\Sigma_3)$ in terms of the other
two. In fact, for the Bianchi I model in the Kasner metric, one of
these quantities is enough to fully represent
$\sigma^{\alpha}\,_{\alpha}$, though for mathematical convenience we
will keep two of these variables by eliminating $\Sigma_1$ in terms
of $(\Sigma_2,\,\Sigma_3)$. By means of (\ref{ec_teta_K}) and
(\ref{sigma_comp}), all second order derivatives of the metric
functions in (\ref{EE2}), (\ref{eq_U[t]}) and (\ref{Maxwell_eq_1})
can be re--written as first order derivatives of $\Theta,\ \Sigma_2$
and $\Sigma_3$. After some algebraic manipulation we can re--write
(\ref{EE2}), (\ref{eq_U[t]}) and (\ref{Maxwell_eq_1}) as the
following autonomous first order system of evolution equations
\begin{eqnarray}\label{EE3}
\dot{U}&=&-(U+p-\frac{2}{3}{\B}{\M})\Theta-\B\M \Sigma_3,
\\
\label{EE31} \dot{\Sigma_2}&=&-\frac{\kappa \B\M}{3}-\Theta\Sigma_2,
\\
\label{EE32}\dot{\Sigma_3}&=&\frac{2}{3}\kappa \B\M- \Theta\Sigma_3,
\\
\label{EE33}\dot{\Theta}&=&\kappa (\B\M+\frac{3}{2}(U-p))-\Theta^2,
\\
\label{EE34}\dot{\beta}&=&\frac{2}{3}\beta (3\Sigma_3-2\Theta).
\end{eqnarray}
Together with these equations, we have the following algebraic
constraint
\begin{equation}\label{constrain}
-\Sigma_2^2-\Sigma_2\Sigma_3+\frac{\Theta^2}{3}-\Sigma_3^2=\kappa U.
\end{equation}
that follows from (\ref{EE2_Gtt}). This non--linear first order
system in the variables $U,\, \beta,\, \Theta,\,
\Sigma_2,\,\Sigma_3$ and the constraint (\ref{constrain}) become
fully determined once we use the thermodynamical equations of the
previous section to express $\M$ and $\B$ in terms of
$\beta=\B/\B_c$. The solution of this system describes the dynamical
evolution of local volumes of a magnetized neutron gas that could be
taken as a rough approximation to a grand canonical subsystem of
this source near the center of a compact object.

\section{Dynamical equations.}

The system of evolution equations (\ref{EE3})--(\ref{EE34}) can be
transformed into a proper dynamical system by introducing the
following dimension--less evolution parameter
\begin{equation}\label{def_tau}
H=\frac{\Theta}{3}, \qquad
\frac{d}{d\tau}=\frac{1}{H_0}\frac{d}{dt},
\end{equation}
together with the dimension--less functions:
\begin{equation}\label{adim_var}
\HH=\frac{H}{H_0}, \ \ S_2=\frac{\Sigma_2}{H_0}, \ \
S_3=\frac{\Sigma_3}{H_0}, \ \ \beta=\frac{\B}{\B_c},
\end{equation}
where $H_0$ is a constant inverse length scale, which for
convenience we choose as $3H_0^2=\kappa\lambda \Rightarrow
|H_0|=1.66\times10^{-4}\,\mathrm{cm^{-1}}$. Note that $H_0$ is not
the cosmological Hubble constant, given by
$H_0^{\rm{cosm}}=0.59\times10^{-28}\mathrm{cm^{-1}}$, but a constant
that provides a length scale $1/H_0\sim 6\,\mathrm{km}$ that is
adequate for the characteristic length scale of the system under
consideration. The functions $S_2$ and $S_3$ are the shear tensor
component normalized to this scale, while the dimension--less time
$\tau$ can be positive or negative, depending on the sign of
$H_0=\pm\sqrt{\kappa\lambda/3}$.

Substituting the variables (\ref{adim_var}) into the system
(\ref{EE3}) and the constraint (\ref{constrain}) we get

\begin{eqnarray}\label{SED}
\mu_{,\tau}&=&\frac{1}{\Gamma_{U,\mu}}\biggl[(2\HH-S_3)(\Gamma_M-2\Gamma_{U,\beta})\beta-3\HH(\Gamma_P+\Gamma_U)
\biggr],
\\
S_{2,\tau}&=&-\beta \Gamma_M-3S_2\HH,
\\
S_{3,\tau}&=&2\beta \Gamma_M-3S_3\HH,
\\
\HH_{,\tau}&=&\beta
\Gamma_M-\frac{3\Gamma_P}{2}-\frac{1}{2}S_2S_3-\frac{3}{2}\HH^2-\frac{1}{2}(S_2^2+S_3^2),
\\
\beta_{,\tau}&=&2\beta(S_3-2\HH),
\end{eqnarray}

\begin{equation}\label{vinculo}
-S_2^2-S_3^2-S_2S_3+3\HH^2=3\Gamma_U.
\end{equation}
where ${}_{,\tau}$ indicates derivative with respect to $\tau$. We
have also replaced the variable $U$ for $\mu$, because from
(\ref{EOS}) we have:\, $U=U(\beta, \mu) \Rightarrow
U_{,\tau}=\lambda(\Gamma_{U,\mu}\mu_{,\tau}+\Gamma_{U,\beta}\beta_{,\tau})$,
which allows us to obtain $\mu_{,\tau}$ from $U_{,\tau}$. We will
solve numerically  the system (\ref{SED}) in the following section.
\section{Numerical solutions and physical discussion }

Since we are interested in studying a collapsing magnetized neutron
gas configuration, we need to consider the local collapse of volume
elements. Hence, we solve the constraint (\ref{vinculo}) to obtain
the two roots of $\HH$, so that the condition for a collapsing
evolution follows by selecting the negative root. To ensure the
local collapse we shall demand in the solution of equations
(\ref{SED}) that the initial expansion, $\Theta$, is negative, which
implies for an initial time hyersurface $\tau=0$ that $\HH(0)<0$.
This follows from (\ref{ec_teta_K}) and (\ref{adim_var}) expressed
in terms of the local proper volume
$V=\sqrt{\det{g_{\alpha\beta}}}=Q_{1}Q_{2}Q_{3}$ as:
\begin{equation}\label{vol_local}
V=V(0)\exp\left(3\int_{\tau=0}^\tau\HH d\tau\right).
\end{equation}
To investigate the direction dependence of the collapse, in terms of
the directions given by the space coordinates $(x,y,z)$, we can
relate by means of (\ref{ec_teta_K}), (\ref{sigma_comp}),
(\ref{def_tau}) and (\ref{adim_var}) the spatial metric components
to the combination $\HH+S_{j}$, leading to:

\begin{equation}
Q_{j}(\tau)=Q_{j}(0)\exp[\int(\HH+S_{j})d\tau], \qquad j=1,2,3.
\end{equation}

where $Q_{j}(0)$, are constants that can be identified with initial
values for $Q_{j}(\tau)$.\\ To solve the system (\ref{SED}) we shall
use a wide range of initial conditions associated with typical
conditions prevailing in a neutron star
\cite{shapiro,bocquet,Peng:2007uu,Reisenegger:2008et,Suh:2000ni,Cardall,salgado1,salgado2},
for example: $\mu=2\Rightarrow \rho\sim10^{15}\,\mathrm{g/cm^3} $,
$\beta_0=10^{-2}-10^{-5}$, for magnetic fields between
$10^{18}\mathrm{gauss}$ and $10^{15}\mathrm{gauss}$. We shall impose
in all numerical trials the condition of volume collapse:
$\HH(0)<0$, together with: $S_2(0)=0,\pm1$, $S_3(0)=0,\pm1$, which
correspond to cases with zero initial deformation and initial
deformation in the direction of the axes $y$ or $z$.\\
The numeric solution for the function $\HH$, displayed in Figure
\ref{Hbeta} for different initial conditions, shows that
$\HH\rightarrow-\infty$, regardless of the selected initial
conditions. The magnetic field tends to increase, but always remains
below the critical field $\B=\B_c$. This behavior is shown in Figure
\ref{Hbeta} for the full range of initial conditions.
\FIGURE{\epsfig{file=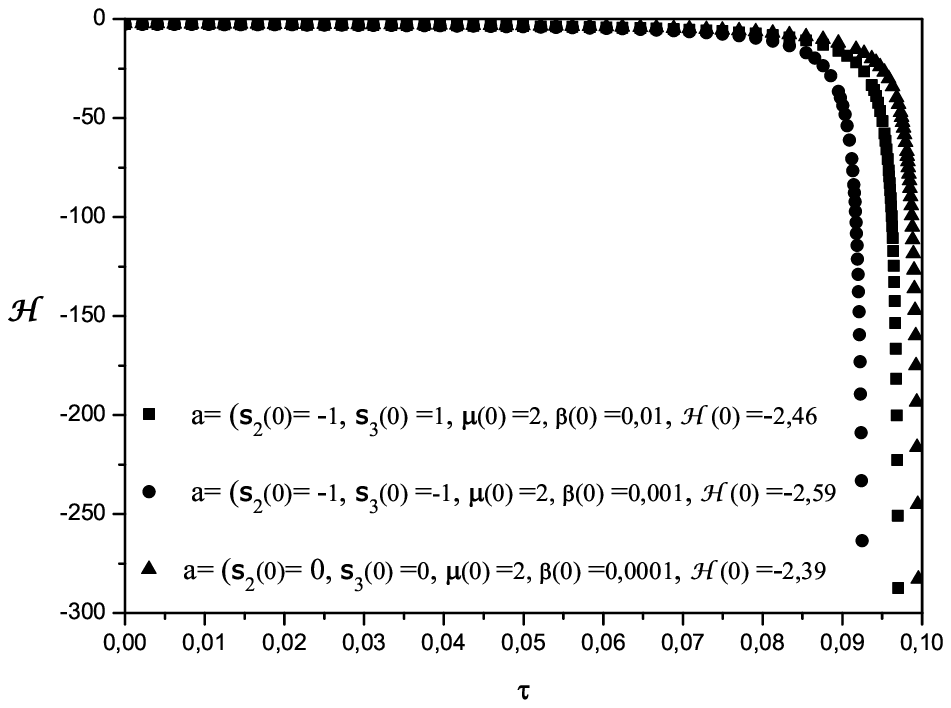,width=7cm}
\epsfig{file=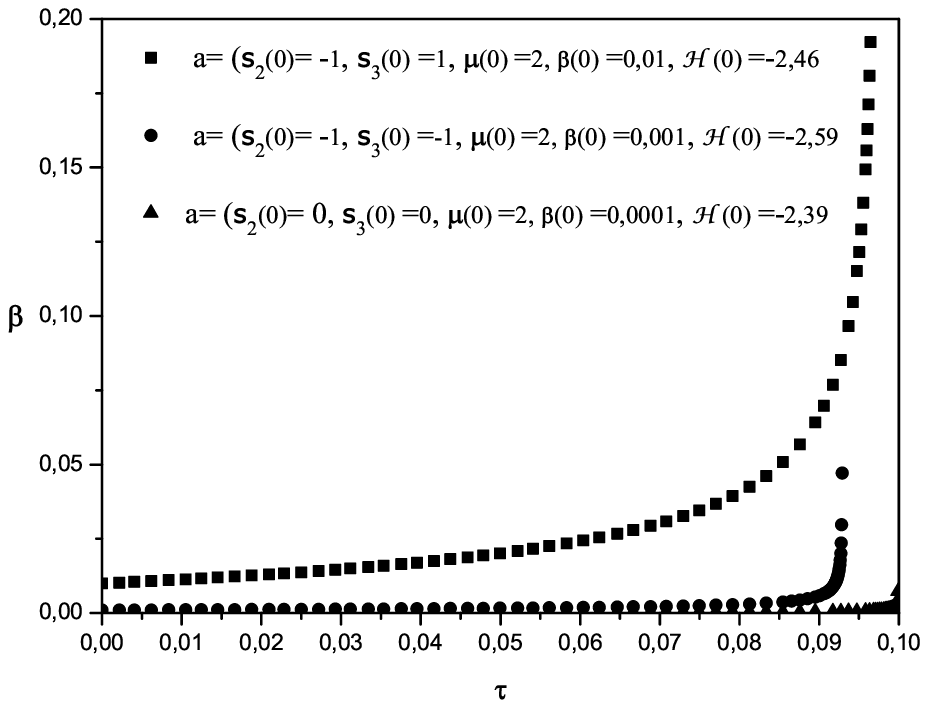,width=7cm} \caption[Example
figure1]{\footnotesize{The left panel shows the behavior of $\HH$ vs
$\tau$ for different initial conditions.The right panel shows the
magnetic field intensity $(\beta=\B/\B_c$), it has tendency to rise,
but remains below the value of the critical field.}} \label{Hbeta}}
The plots displayed in Figure \ref{S1S2} and in the left panel of
Figure \ref{S3espfase} describe the collapse of the fluid elements
from the solutions of (\ref{SED}).
\FIGURE{\epsfig{file=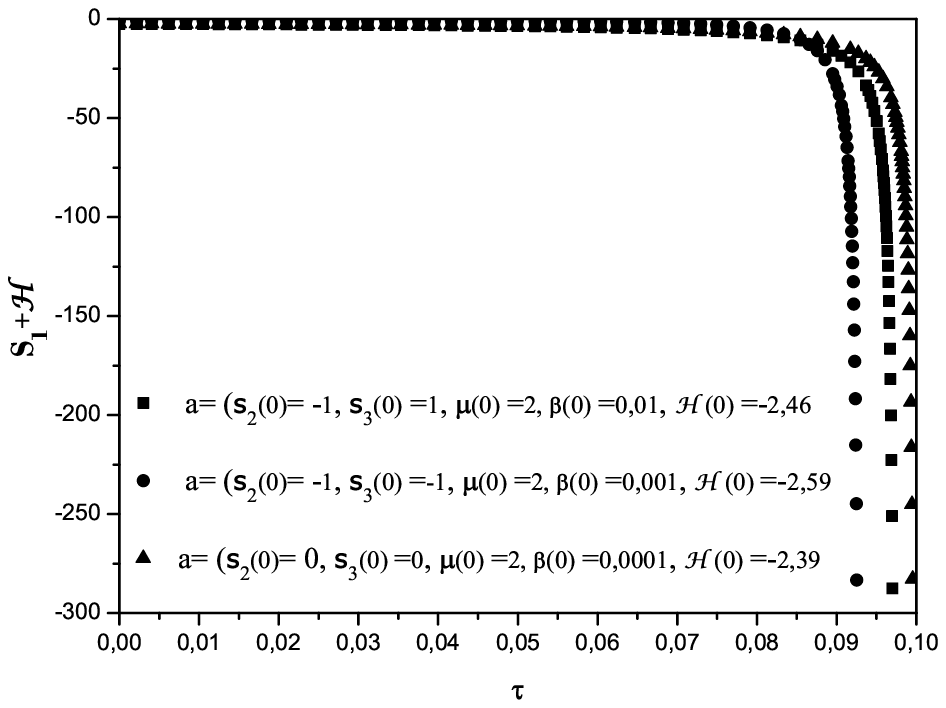,width=7cm}
\epsfig{file=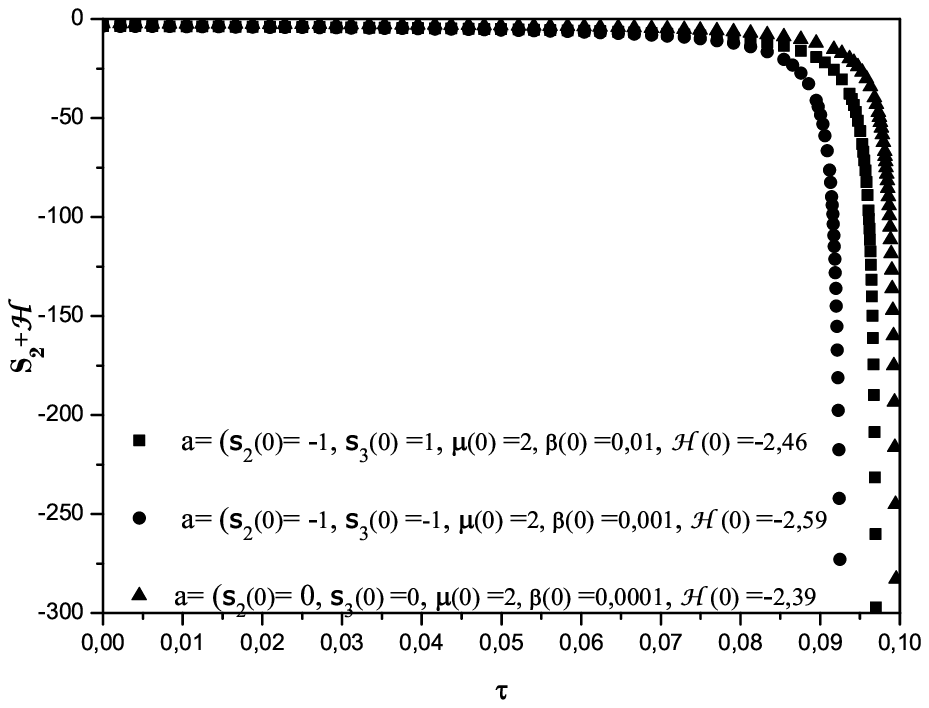,width=7cm} \caption[Example
figure2]{\footnotesize{Behavior of $(S_1+\HH)$ and $(S_2+\HH)$
versus $\tau$. We can see this quantities tending to $-\infty$,
besides different collapse times for different initial conditions.}}
\label{S1S2}}
\FIGURE{\epsfig{file=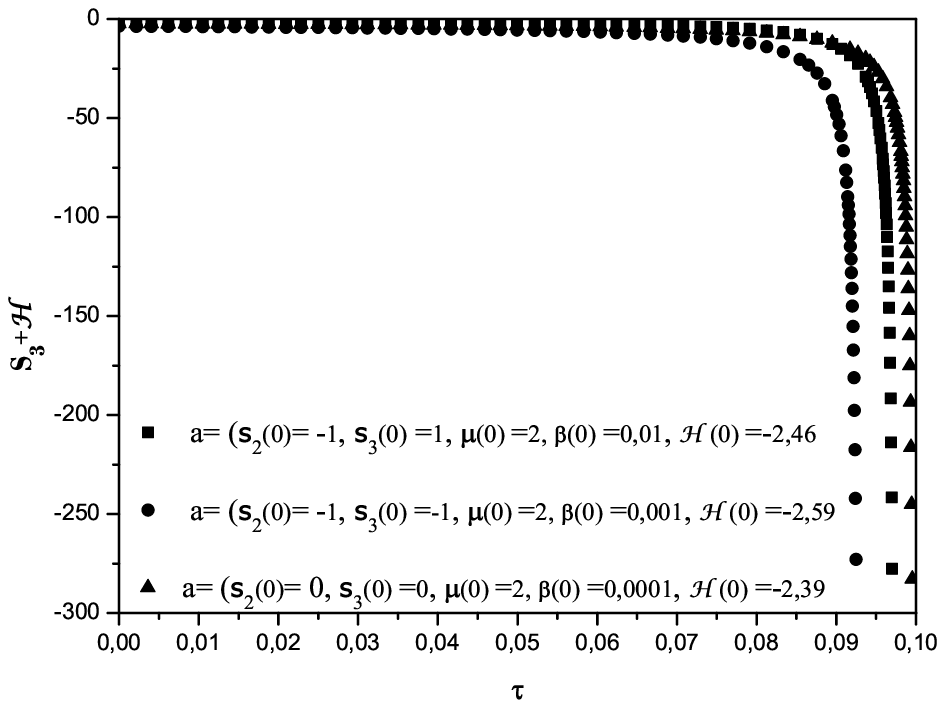,width=7cm}
\epsfig{file=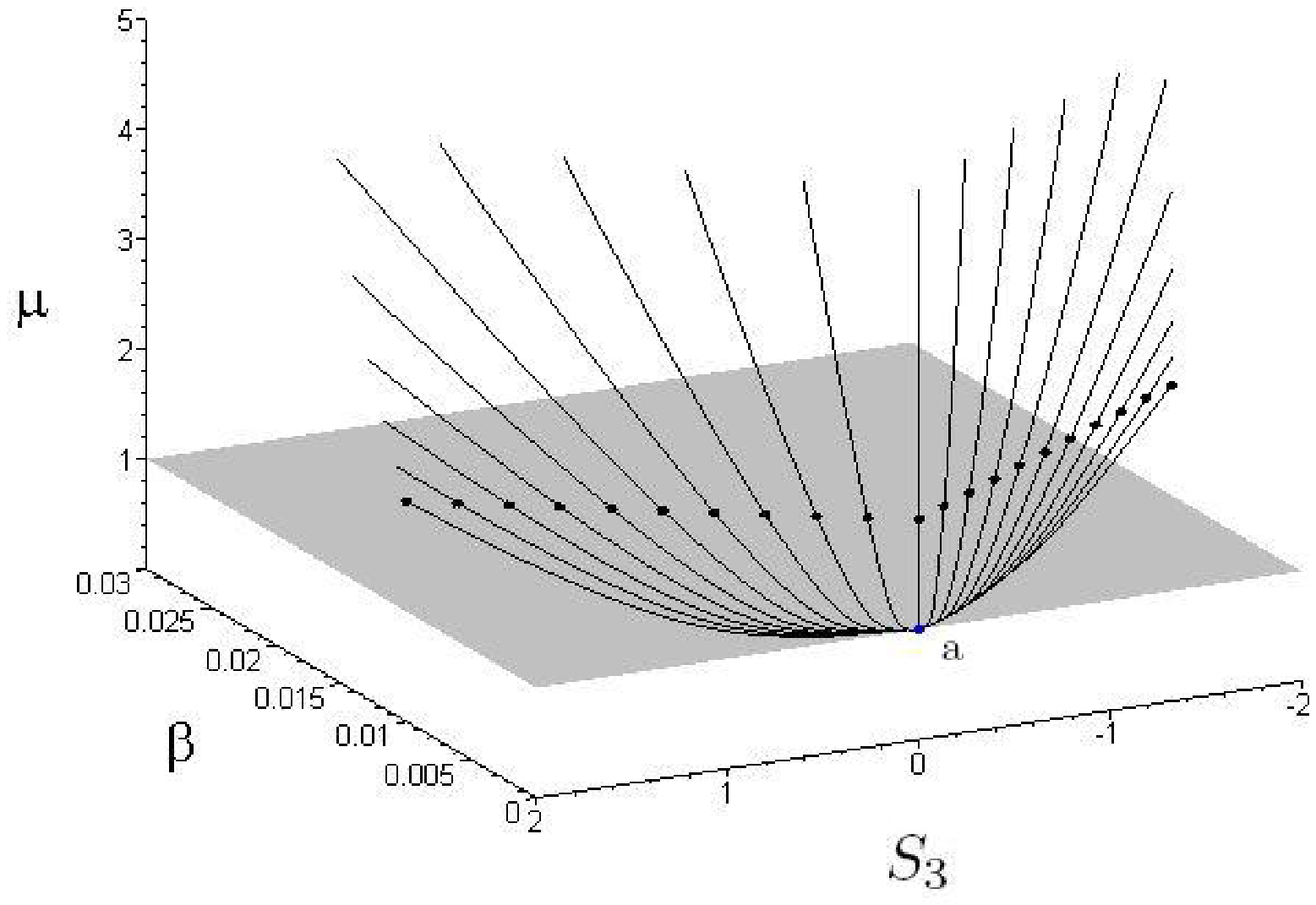,width=7cm} \caption[Example
figure3]{\footnotesize{The left panel shows behavior of $(S_3+\HH)$
versus $\tau$. The behavior is similar to the trajectories given in
Fig\,\ref{S1S2}, therefore the quantity $(S_3+\HH) \rightarrow
-\infty$. The right panel shows paths in the small section of the
space phase $(S_3, \beta, \mu)$. The points represent the unbranded
initial conditions, the point \textquotedblleft a\textquotedblright\
represents the attractor point which coordinates are
$(S_3=0,\beta=0,\mu=1,\HH=0)$.}} \label{S3espfase}}
It is evident from these figures that the quantities
$S_i+\HH\,\rightarrow\,-\infty$, so that the spatial metric
coefficients tend to zero ($Q_{1}, Q_{2}, Q_{3}\,\rightarrow0$),
which clearly shows that volume elements collapse to a point like
isotropic singularity.
\section{Phase Space.} As it was done in \cite{Alain2}, we can use
the constraint (\ref{vinculo}) to transform the evolution system
(\ref{SED}) to a reduced system of equations in the variables $S_3,
\beta, \mu, \HH$

\begin{eqnarray}\label{SED_2}
\mu_{,\tau}&=&\frac{1}{\Gamma_{U,\mu}}\biggl[(2\HH-S_3)(\Gamma_M-2\Gamma_{U,\beta})\beta-3\HH(\Gamma_P+\Gamma_U)
\biggr],
\\
S_{3,\tau}&=&2\beta \Gamma_M-3S_3\HH,
\\
\HH{,\tau}&=&\beta \Gamma_M+\frac{3}{2}(\Gamma_U-\Gamma_P)-3\HH^2,
\\
\beta_{,\tau}&=&2\beta(S_3-2\HH),
\end{eqnarray}
where we note that the only equation that was modified is the
equation for $\HH$, hence (\ref{SED_2}) is equivalent to
(\ref{SED}).

 The trajectories in the 3--dimensional subsection of the phase space,
 parametrized by $(S_3, \beta, \mu)$,
are shown in Figure \ref{S3espfase}. The evolution of the system is
determined by the sign of $H_0$. For $\tau<0 \Rightarrow
H_0=-\sqrt{\kappa\lambda/3}$, the system evolves towards the stable
attractor (point marked by a), while for $\tau>0 \Rightarrow
H_0=\sqrt{\kappa\lambda/3}$, the trajectories evolve towards a
singularity. A similar study was conducted for the remaining
3--dimensional subsections of the space phase, obtaining
qualitatively similar results. which was obtained coordinates of
attractor are: $(S_3=0, \beta=0, \mu=1, \HH=0)$.
\section{Conclusions.}
We have used a Bianchi I model to study the evolution of a
magnetized neutron gas characterized by a physically motivated and
fully relativistic equation of state.  As far as we are aware, this
equation of state has not been considered previously in a general
relativistic context. Besides the general theoretical interest in
undertaking such a study, we argue that the simplified Bianchi
geometry roughly approximates a grand canonical subsystem of a
magnetized neutron source in the conditions prevailing near the
center of a compact object, hence our treatment can be conceived as
a toy model that can be useful in understanding the local evolution
of volume elements of this source under these conditions. However, a
proper examination of the specific limitations of the dynamics of
this toy model and/or its connection with concrete astrophysical
studies of actual compact objects lies beyond the scope of the
present article. As we comment further ahead, we will consider these
important tasks in forthcoming articles by resorting to perturbation
techniques, more elaborate numerical methods and less idealized
sources.

The Einstein-Maxwell field equations for a magnetized neutron gas in
the Bianchi I geometry were transformed into a set of non--linear
evolution equations, which were solved numerically for generic
collapsing initial conditions and analyzed qualitatively as a proper
dynamical system. The results that we found are:
\begin{itemize}
\item The final state in the collapsing evolution of local volume elements is an isotropic point--like singularity.
This final state occurred for a wide range of initial conditions
associated with parameter values that would be typical in a compact
object.
\item The magnetic field increases, but its values are always below the critical field.
This result is consistent with numerical values of maximal field
intensities compatible with stability in numerical studies of
magnetized rotating configurations (see \cite{bocquet}).
\item The study of the phase space associated with the dynamical equations shows
that the system evolves, for $H_0<0$, to an equilibrium point, ({\it
i.e.} into a stable configuration). It is possible to introduce a
temperature dependence in the equation of state. Buy doing so, the
of evolution of the neutron gas could be associated with high
temperature neutron sources in the context of early universe
conditions in cosmological models dealing with primordial magnetic
field \cite{King:2006cy}.
\end{itemize}

It is important to remark that, unlike the dynamical study of a
magnetized electron gas~\cite{Alain2}, anisotropic ``cigar--like''
singularities did not occur for a wide range of initial conditions.
Since a dynamical effect of the magnetic field under critical
conditions is a final state anisotropic singularity aligned in the
direction of the field, the exclusive occurrence of point--like
isotropic in a magnetized neutron gas suggests that the final stage
of the evolution of this gas is more intensely dominated by the
focusing effect associated with strong gravity than the electron
gas. This is consistent with the fact that electrons are strongly
coupled to the magnetic field through their electric charge, whereas
neutrons have a weaker coupling associated with their anomalous
magnetic moment.

 It is important to stress that the geometry of Bianchi I models in a
comoving frame has stringent limitations in dealing adequately with
the dynamical effects associated with a magnetic field. This is
important when considering neutrons as a source in which electric
charge vanishes but not the magnetic moment. By being spatially
homogeneous with a 4--velocity orthogonal to flat 3--dimensional
hypersurfaces of maximal symmetry, the Lorentz force is necessarily
zero: $f^{b}=qu_{a}F^{a b}=0$. Also, the fact that the Bianchi I
model is spatially flat makes it inadequate to examine (even as a
toy model) the interplay of local collapse and the magnetic field in
localized objects, as such interplay is necessarily associated with
strong positive spatial curvature. However, in our case these
inadequacy can be overcome (at least partially) by considering a
general perturbation scheme on a Bianchi I model, in which spatial
curvature and 4-acceleration are perturbative but not zero. In this
case it is possible to examine the effects of the magneto--curvature
coupling associated with a non--trivial Lorentz force and a nonzero
deceleration parameter in the Raychaudhuri equation (see
\cite{Tsagas} for general detail). The use of such a perturbed
Bianchi I model for the description of the neutron gas source
considered in this article is presently under consideration in a
separate article.

Besides the introduction of a perturbation scheme in a Bianchi I
model, another possible improvement on the dynamical description of
the source under consideration would be to consider Bianchi models
I, V, VII or IX with a tilted 4--velocity, which are endowed with
more degrees of dynamical freedom, including even the possibility of
nonzero rotation (see \cite{Coley:2008gh} and references quoted
therein). These models would allow for a less restrictive study of
the dynamical effects, reported in \cite{Tsagas}, in which magnetic
tension and gravitational collapse may present non--trivial coupling
with a nonzero and non--perturbative 4--acceleration and vorticity
with the magnetic field.

Finally, as we mentioned in the introduction, a magnetized gas
consisting only of neutrons can be theoretically interesting but it
is too idealized as a potential source for a compact object. Thus,
we will consider as an extension of this work, besides the extra
degrees of freedom in the dynamics (mentioned above), a gas mixture
of neutrons, electrons and protons, complying with suitable balance
conditions and adequate chemical potentials, in comparison with
other types of equations of state \cite{bocquet,salgado1,salgado2}.
These extension of the present work are also under consideration for
future articles.
\section*{Acknowledgements}
{The work of A.P.M and A.U.R has been supported by \emph{Ministerio
de Ciencia, Tecnolog\'{\i}a y Medio Ambiente} under the grant CB0407
and the ICTP Office of External Activities through NET-35.  A.P.M.
also acknowledges the Program of Associateship TWAS-UNESCO-CNPq as
well as the hospitality and support of CBPF through the program
PCI-MCT. R.A.S. and A.U.R. acknowledge support from the research
grant \emph{DGAPA--UNAM PAPIIT--IN119309}.}
%
%

%
\end{document}